\documentstyle[aps,prd,preprint]{revtex}

\newcommand{\beq}{\begin{equation}}
\newcommand{\eeq}{\end{equation}}
\def\beqa{\begin{eqnarray}}
\def\eeqa{\end{eqnarray}}

\def\nn{\nonumber}

\begin{document}

\title{Gravity of higher-dimensional global defects}

\author{Itsaso Olasagasti and Alexander Vilenkin}

\address{Institute of Cosmology,
Physics Department, Tufts University, Medford, MA 02155, USA}

\maketitle

\begin{abstract}
Solutions of Einstein's equations are found for global defects in a
higher-dimensional spacetime with a nonzero cosmological constant
$\Lambda$.  The defect has a $(p-1)$-dimensional core (brane) and a
`hedgehog' scalar field configuration in the $n$ extra dimensions.
For $\Lambda = 0$ and $n > 2$, the solutions are characterized by a
flat brane worldsheet and a solid angle deficit in the extra
dimensions.  For $\Lambda > 0$, one class of solutions describes
spherical branes in an inflating higher-dimensional universe.
Instantons obtained by a Euclidean continuation of such solutions
describe quantum nucleation of the entire inflating brane-world, or of
a spherical brane in an inflating higher-dimensional universe.  For
$\Lambda < 0$, one class of solutions exhibits an exponential warp
factor.  It is similar to spacetimes previously discussed by Randall
and Sundrum for $n = 1$ and by Gregory for $n = 2$.
\end{abstract}

\section{Introduction}

In recent years there has been renewed interest in `brane-world'
models in which the universe is represented by a $(3+1)$-dimensional
subspace (3-brane) embedded in a higher-dimensional (bulk) spacetime
\cite{Gia,RS}.  In such models, all the familiar matter fields are
constrained to live on the brane, while gravity is free to propagate
in the extra dimensions.  Initially it was thought
that realistic models required compact extra dimensions, but it 
has been shown by Randall and Sundrum \cite{RS} that it is possible to have
infinite extra dimensions and still have gravity effectively localized
on the brane.  This is achieved by introducing a negative cosmological
constant which has the effect of `warping' the extra-dimensional
space, so that most of the physical volume is concentrated near the
brane. 

In most of the recent work, including that of Randall and Sundrum, the
brane is pictured as a domain wall propagating in a 5-dimensional bulk
spacetime.  The case of two extra dimensions has also been considered,
when the branes are similar to strings and the bulk has 6 dimensions.
For a gauge string, the metric outside the string core is flat with a
conical deficit angle, and Sundrum \cite{Sundrum} suggested that the
extra dimensions can be compactified by introducing a sufficient
number of branes, so that the total deficit angle is equal to $2\pi$. 
This was generalized by Chodos and Poppitz \cite{CP} to include a
positive cosmological constant.  Cohen and Kaplan \cite{CK} considered
the case of a global sting which has a curvature singularity at a
finite distance from the string core.  They argued that the
singularity can provide an effective compactification of the extra
dimensions.  Gregory \cite{Gregory} has shown that a non-singular
global string solution exists in the presence of a negative
cosmological constant.  In this solution the extra dimensions are
infinite and strongly warped, as in the Randall-Sundrum model.
Garriga and Sasaki \cite{GS} discussed a Euclidean continuation of the
Randall-Sundrum spacetime and interpreted the resulting instanton as
describing quantum nucleation of a 5-dimensional brane-world from
nothing. 

In this paper, we shall explore a more general case of a brane carrying
a global charge in a higher-dimensional spacetime with a nonzero
cosmological constant $\Lambda$.  
We shall consider a $(p-1)$-dimensional brane in a bulk spacetime of
$D=p+n$ dimensions.  The physically interesting case is $p=4$, but we
shall allow an arbitrary $p$ for greater generality.
For $n=3$ and $\Lambda=0$, one can expect to
recover the global monopole metric with a solid deficit angle
\cite{BV}, but for $n>3$ the defects do not have 3-dimensional
analogues.  

The paper is organized as follows.  In the next Section we introduce
the scalar field and metric {\it ansatz} and present the corresponding
Einstein's equations.  Our solutions for $\Lambda\geq 0$ and $\Lambda<
0$ are given, respectively, in Sections III and IV.  Euclidean
instanton solutions are discussed in Section V, and our conclusions
are summarized in Section VI.

\section{Einstein's equations}

We shall use the notation
$\{x^{\mu}\}$ with $\mu=0, ... ,p-1$ for the coordinates on the brane
worldsheet,  
$\{\xi^{a}\}$ with $a=1, ... ,n$ for coordinates in the extra
dimensions, and 
$\{X^{A}\}$ with $A=0, ... ,D-1$ for general coordinates in the 
$D$-dimensional spacetime.
 
A global defect in $n$ extra dimensions is described by a multiplet of
$n$ scalar fields $\phi^a$ with a Lagrangian
\beq
L={1 \over 2}\; \partial_A \phi^a \partial^A \phi^a
-V(\phi),
\eeq
where the potential $V(\phi)$ has its minimum on the $n$-sphere
$\phi^a\phi^a=\eta^2$.  One can use, for example,
\beq
V(\phi)={\lambda\over{4}}(\phi^a\phi^a-\eta^2)^2.
\eeq
The defect solution should have $\phi=0$ at the center of the defect
and approach the radial `hedgehog' configuration outside the core,
\beq
\phi^a(\xi)=\eta\;{\xi^a \over \xi}
\label{fieldc}
\eeq
with $\xi^2\equiv \xi^a \xi^a$.  We shall be interested only in the
exterior solutions, where $V(\phi)\approx 0$ and $\phi(\xi)$ is
accurately approximated by (\ref{fieldc}). 

We shall adopt the following {\it ansatz} for the metric:
\beq
ds^2=A(\xi)^2 d\xi^2+\xi^2 d\Omega^2_{n-1}
+B(\xi)^2 {\hat g}_{\mu \nu} dx^{\mu} dx^{\nu},
\label{ansatz}
\eeq
where $d\Omega^2_{m}$ stands for the metric on a unit $m$-sphere, and
the spherical coordinates in the extra dimensions are defined by the
usual relations, $\xi^a=\{\xi \cos\theta_1, ~\xi \sin\theta_1\cos\theta_2,
... \}$. (A different {\em ansatz} will be considered in sections III.C
and IV.C). 
The energy-momentum tensor for the field configuration
(\ref{fieldc}) is then given by
\beqa
& T^{\xi}_{\xi}=-{1\over 2}\; (n-1) \; {\eta^2 \over 
\xi^2},& \nn \\
& T^{\theta_a}_{\theta_b}=-{1\over 2}\; (n-3) \; {\eta^2 \over \xi^2}
\; \delta^a_b,& \nn \\
& T^{\nu}_{\mu}=-{1\over 2}\; (n-1) \; {\eta^2 \over 
\xi^2} \; \delta_\mu^\nu . \nn \\&
\label{TAB}
\eeqa

Our goal will be to solve Einstein's equations in $D$ dimensions,
\beq
R_{AB}-{1 \over 2}\; G_{AB} R= \kappa^2 T_{AB}- \Lambda G_{AB},
\eeq
where $G_{AB}$ is the $D$-dimensional metric, 
$\Lambda$ is the cosmological constant and $T_{AB}$ is from
Eq.(\ref{TAB}).  
The line element (\ref{ansatz}) is a special case of the more general class of 
metrics,
\beq
ds^2= d\tilde{s}_n^2+B(\xi^a)^2 d\hat{s}^2,
\label{gen}
\eeq
where $d\tilde{s}_n$ depends only on the transverse coordinates
$\{\xi^a\}$ and $d\hat{s}^2$ only on those on the brane $\{x^{\mu}\}$.
For such metrics, the Ricci tensor splits in the following way:
\beqa
R_{m n}&=&\tilde{R}_{m n}-p{B_{;mn}\over B}, \label{ricci1}\\
 R_{\mu \nu}&=&\hat{R}_{\mu \nu} -\hat{g}_{\mu \nu}
\left[ B \tilde{\nabla}^2 B+(p-1)(\tilde{\nabla} B)^2 \right].
 \label{ricci2}
\eeqa

Since  $T_{\mu \nu} \propto {\hat g}_{\mu \nu}$, through Einstein's
equations we  
have that $R_{\mu \nu} \propto {\hat g}_{\mu \nu}$ and finally, from equation 
(\ref{ricci2}) above, that $\hat{R}_{\mu \nu} \propto {\hat g}_{\mu \nu}$.
That is, $\hat{R}$, the curvature associated with the metric
${\hat g}_{\mu \nu}$, must be constant. 
Einstein's field equations then reduce to
\beqa
{1\over A^2}
\left[ p(p-1)\;\left({B'\over B }\right)^2 +2p\;{n-1\over \xi}\;{B'\over B 
} +{(n-1)(n-2)\over \xi^2} \right]
 \nn \\
-{n-1\over \xi^2}
(n-2-\kappa^2 \eta^2)+2\Lambda-{\hat{R}\over B^2 }& =0&,
\label{efe1} \\
 {1 \over A^2}
\left[ 2p\left({B'' \over B}-{A' \over A}{B'\over B}\right)
+p(p-1)\left({B' \over B}\right)^2+
{2(n-2)\over \xi}\left(p{B'\over B}-{A' \over A}\right)
\right.
&& \nn \\
\left.
+{(n-3)(n-2)\over  
\xi^2} \right] 
-{n-3\over \xi^2}(n-2-\kappa^2\eta^2) +2\Lambda
-{\hat{R}\over B^2}&=0, &\\
{1 \over A^2}
\left[ 2(p-1) \left({B'' \over B} -{A' \over A}{B'\over B}\right)
+{2(n-1)\over \xi}\left((p-1){B'\over B}-{A'\over A} \right) 
+(p-1)(p-2)\left({B'\over B }\right)^2 
\right.
&& \nn \\ 
\left.
+{(n-1)(n-2)\over \xi^2} 
\right] 
-{n-1\over \xi^2}(n-2-\kappa^2\eta^2)+2\Lambda
+ {2-p\over p}{\hat{R}\over B^2}&=&0,
\label{efe3}
\eeqa
supplemented by the equation for the metric on the $(p-1)$-brane,
\beq
\hat{R}_{\mu \nu}={\hat{R}\over p}\; {\hat g}_{\mu \nu}.
\label{bm}
\eeq
It can be shown that only two of the three equations
(\ref{efe1})-(\ref{efe3}) are independent.

We have been able to find several classes of solutions to this set of
equations.  We shall discuss separately the cases of positive and
negative $\Lambda$ and consider both Lorentzian and Euclidean versions
of the metric.

\section{Solutions with $\Lambda\geq 0$}

\subsection{Class I}

The first class of solutions is obtained with the
{\em ansatz} $A(\xi)=B(\xi)^{-1}$, which is the same as the one used
for a global monopole in \cite{BV}.  With this {\em ansatz},
Einstein's equations are considerably simplified. The two independent 
equations can be written as
\beq
(p+1){A'\over A^3}\xi-{n-2 \over A^2}+(n-2)-\kappa^2\eta^2-
{2 \Lambda \over n+p-2}\xi^2 =0,
\label{EE1}
\eeq
\beq
-{A''\over A^3}+(p+2) \left( {A'\over A^2}\right)^2
-{(n-1)\over \xi} {A'\over A^3}
-{\hat{R}\over p }A^2+{2 \Lambda \over n+p-2}=0,
\label{EE2}
\eeq
and we find the following solution:
\beqa
& A^{-2}(\xi)=B^2(\xi)=1-{\kappa^2 \eta^2 \over n-2}\; 
            -{2 \Lambda \over (n+p-2)(n+p-1)}\; \xi^2 &
      \label{s11} \\
& \hat{R}={2\Lambda p (p-1)\over (n+p-1)(n+p-2)}
\; (1-{\kappa^2 \eta^2\over n-2})& \label{s12}
\eeqa

The solution (\ref{s11}) is only valid for $n>2$. For $n=3$ and
$\Lambda=0$, the transverse to the brane part of the solution
coincides with the metric of a global monopole.  In higher dimensions,
$n>3$, the form of the metric is quite similar: the defect
introduces a solid angle deficit in extra dimensions.  This is
remarkable, considering the fact that defect solutions are quite
different for $n=1,2$.
  
With an appropriate rescaling of the coordinates, the $\Lambda=0,
~n\geq 3$ solution can be written as
\beq
ds^2=d\xi^2+\left(1-{\kappa^2\eta^2\over{n-2}}\right)\xi^2 d\Omega^2_{n-1}
+\eta_{\mu \nu} dx^{\mu} dx^{\nu},
\eeq
where $\eta_{\mu\nu}$ is the Minkowski metric.  As the symmetry
breaking scale $\eta$ is increased, the solid angle deficit grows and
eventually consumes the entire solid angle at the critical value
\beq
\eta_c =(n-2)^{1/2}\kappa^{-1}.  
\label{etac}
\eeq
One expects that the transverse dimensions in this case have the
geometry of an infinite cylinder whose cross-sections are
$(n-1)$-spheres of a fixed radius.  This expectation will be verified
in Section III.C.

We next consider solutions with $\Lambda>0$.
Requiring that the right-hand side of (\ref{s11})
is positive, we should have
$\kappa^2\eta^2/(n-2)<1$, so that $\hat{R}>0$, and
$\xi$ should be constrained to the
interval $0<\xi<\xi_m$ where $\xi_m$, defined by the condition
$B(\xi_m)=0$, is
\beq
\xi_m^2={(n+p-2)(n+p-1)\over 2\Lambda}\;\left( 
1-{\kappa^2 \eta^2\over n-2}
  \right).
\eeq

Another form of the metric can be obtained using the transformation:
\beq
\xi=\xi_m \sin \chi,
\eeq
which gives
\beq
ds^2=K \left[ d\chi^2+ 
\alpha^2\sin^2\chi
d\Omega_{n-1}^2+\cos^2 \chi d\hat{s}_+^2 
\right], \label{c11}\\
\eeq
\beq
{\hat R}=p(p-1),
\label{Rp}
\eeq
where
\beq
K=(n+p-1)(n+p-2)/2|\Lambda|,
\label{K}
\eeq
\beq
\alpha^2=\left|1-{\kappa^2\eta^2\over{n-2}}\right|,
\label{alpha}
\eeq
and $\chi$ varies in the interval $0<\chi<\pi$.  The absolute value
signs on the right-hand sides of Eqs.(\ref{K}),(\ref{alpha}) are
introduced for later use.

The positive-curvature metric $d\hat{s}_+^2$ can be given by any solution of
Eq.(\ref{bm}) with ${\hat R}$ from Eq.(\ref{Rp}).  In this paper we
shall assume it to be the $p$-dimensional de Sitter space, 
\beq
d\hat{s}_+^2=
     (-dt^2+\cosh^2 t \;d\Omega_{p-1}^2).
\label{ds1}
\eeq
This is the case of highest symmetry, when all points on the brane
worldsheet are equivalent.

The transverse part of the solution (\ref{c11}) describes an $n$-sphere
with a solid angle deficit.  This may create an impression that the
extra dimensions are compactified with a fixed compactification radius
$\sim \sqrt{K}$.  However, this impression is misleading.  
In the limit $\eta\to 0$, the deficit angle vanishes, and the
metric (\ref{c11}),(\ref{ds1}) becomes that of $(p+n)$-dimensional de
Sitter space written in somewhat unfamiliar coordinates \cite{Rama}.  
With a more
familiar form of de Sitter metric,
\beq
ds^2=K(-dt^2 +\cosh^2t ~d\Omega_{p+n-1}^2),
\label{exp}
\eeq
it is clear that all dimensions are equally large and expanding.  
Spatial sections of the universe are $(p+n-1)$-spheres, and spatial
sections of the brane are $(p-1)$-spheres of the same radius.  So the
brane is wrapped
around the universe along one of the `big circles'.  Both the brane
and the universe expand exponentially with time.  A
nonzero $\eta$ introduces a deficit angle, but does not change the
qualitative character of the spacetime.  

We note that de Sitter space also appears to be static in the coordinates
\beq
ds^2=K(-\cos^2\psi ~dt^2 +d\psi^2+\sin^2\psi ~d\Omega_{p+n-2}^2).
\label{static}
\eeq
The reason for this is well known: this coordinate system does not
cover the whole spacetime; it covers only the interior of a sphere of
radius equal to the de Sitter horizon.  Our solution
(\ref{c11}),(\ref{ds1}) uses a mixed representation in which the
metric has a static form like (\ref{static}) in the transverse dimensions 
and an expanding form like (\ref{exp}) on the brane.  The coordinate
system in (\ref{c11}) covers the region from $\chi=0$ to the horizon
surface $\chi=\pi/2$ where the determinant of the metric vanishes,
indicating a coordinate singularity.

The metric (\ref{c11}) is somewhat similar to the dilatonic string
solution found by Dando and Gregory \cite{DG}. They interpreted their
solution as describing a string-antistring pair in a universe with
compact static transverse dimensions. Our interpretation of (\ref{c11}) 
is quite different, and we believe a similar interpretation should
also apply to the Dando - Gregory solution.

The induced metric on the brane is $ds_p^2=K d\hat{s}^2$, and the
curvature of the brane worldsheet is
\beq
R_p={2 \Lambda p(p-1) \over (n+p-1)(n+p-2)}.
\eeq
This shows that the curvature of the
brane is determined only by the cosmological constant 
$\Lambda$, while the symmetry breaking scale $\eta$ affects only the
deficit angle in the extra dimensions.  

The solution (\ref{c11}) has curvature singularities at $\chi=0,\pi$
(since $T_A^B$ is singular there), but these singularities are rather
mild, and the metric coefficients are non-singular.  One should
remember that Eq.(\ref{c11}) gives a solution only in the exterior
region outside the defect core.  But since the metric is well-behaved
at $\chi\to 0,\pi$, one can expect that it gives a reasonably accurate
representation of the full spacetime in the limit when the defect
thickness $\delta$ can be neglected, $\delta\ll \xi_m$.

\subsection{Class II}

We find another solution to the equations 
(\ref{efe1})-(\ref{efe3})
by considering a different {\em ansatz}: $B(\xi)=\xi$. Again the equations
simplify considerably and it is possible to find an analytic solution:
\beqa
& A^{-2}(\xi)={n-2-\kappa^2 \eta^2 \over n+p-2}\; 
            -{2 \Lambda \over (n+p-2)(n+p-1)}\; \xi^2, \quad 
B(\xi)=\xi & \label{s21}\\
& \hat{R}=p\; (n-2-\kappa^2 \eta^2)& \label{s22} 
\eeqa
As in the previous case, for $\Lambda>0$ we have from the condition
$A^2>0$ that $n-2-\kappa^2 \eta^2>0$, and thus 
$\hat{R}>0$ and $\xi$ is constrained to the interval
$0<\xi<\xi_m$ with
\beq
\xi_m^2={(n+p-1)(n-2-\kappa^2\eta^2)\over 2\Lambda}
\eeq

As before, we redefine the radial coordinate as
\beq
\xi=\xi_m \sin \chi,
\eeq
and the metric takes the form
\beq
ds^2=K \left[ d\chi^2+ {\tilde\alpha}^2 \sin^2 \chi
(d\Omega_{n-1}^2+d\hat{s}^2) 
\right] 
\label{class2}
\eeq
where $K$ is given by Eq.(\ref{K}), 
\beq
{\tilde\alpha}^2=\left|{n-2-\kappa^2 \eta^2 \over n+p-2}\right|,
\eeq
$d\hat{s}^2$ stands for a $p$-dimensional spacetime of constant curvature 
$\hat{R}=p(p-1)$, and $\chi$ takes values in the interval $0<\chi<\pi$. 

An unphysical feature of the solution (\ref{class2}) is that the
deficit angle does not vanish even for $\eta=0$, that is, in the
absence of a defect.  We have verified that the curvature invariant
$R^{\mu\nu\sigma\tau}R_{\mu\nu\sigma\tau}$ diverges at $\chi=0,\pi$
for $\eta=0$.  These singularities appear to be unrelated to the
defect, and we dismiss class-II solutions as unphysical.

\subsection{Class III}

As we mentioned in Section III.A, the `conical' geometry of the extra
dimensions is expected to degenerate into a cylinder at some critical
value of the symmetry breaking scale $\eta$.  In order to verify this
expectation, we introduce the following {\it ansatz}:
\beq
ds^2=d\xi^2+C^2 d\Omega_{n-1}^2+B(\xi)^2 d\hat{s}^2,
\label{newansatz}
\eeq
where $C$ is a constant radius of the $(n-1)$-spheres.
This is again of the form (\ref{gen}), so Eqs.(\ref{ricci1}), 
(\ref{ricci2}) can be used, and we obtain
\beqa
-p {B''\over B}-{2\Lambda\over n+p-2} &=&0, \label{cyl1}\\
{n-2-\kappa^2\eta^2\over C^2}-{2\Lambda\over n+p-2} &=&0, \label{cyl2}\\
{1 \over p} {\hat{R}\over B^2}-{B''\over B}
-(p-1)\left({B'\over B} \right)^2-{2\Lambda\over n+p-2} &=&0. \label{cyl3}
\eeqa

For $\Lambda=0$, Eq.(\ref{cyl2}) gives $\eta=(n-2)^{1/2}\kappa^{-1}$,
which agrees with the critical value (\ref{etac}).  From
Eq.(\ref{cyl1}), $B'={\rm const}$, and it follows from (\ref{cyl3})
that the worldsheet
curvature ${\hat R}$ can be either positive or zero.
For ${\hat R}=0$, $B={\rm const}$, and the solution is 
\beq
ds^2=d\xi^2+C^2 d\Omega^2_{n-1}
+\eta_{\mu \nu} dx^{\mu} dx^{\nu}.
\label{cigar}
\eeq
The radius of the cylinder $C$ is arbitrary; we expect it to be
determined by matching to an appropriate interior solution in the
defect core, with the complete geometry being that of a `cigar'.

For ${\hat R}>0$ and with a suitable normalization of the radial
coordinate, the
solution can be written as
\beq
ds^2=C^2d\Omega_{n-1}^2 +d\chi^2+\chi^2 d{\hat s}_+^2
\label{rt}
\eeq
with $d{\hat s}_+^2$ from Eq.(\ref{ds1}).  It can be shown
that the last two terms in the metric (\ref{rt}) describe a
$(p+1)$-dimensional
Minkowski space in unfamiliar coordinates \cite{RT}.  This metric is
therefore equivalent to (\ref{cigar}).

For $\Lambda>0$, Eq.(\ref{cyl2}) gives
\beq
C^2=(n+p-2)(n-2-\kappa^2\eta^2)/2\Lambda,
\eeq
and we find a solution of the form
\beq
ds^2=C^2d\Omega_{n-1}^2 +\omega^{-2}(d\chi^2+\sin^2\chi d{\hat s}_+^2),
\label{rtds}
\eeq
where
\beq
\omega=\sqrt{{2\Lambda\over p(n+p-2)}}.
\eeq
The last two terms in the metric (\ref{rtds}) describe a $(p+1)$-dimensional
de Sitter space. Note that, in contrast to the $\Lambda=0$ case, 
solutions now exist for all values of $\eta<\eta_c$, while for $\eta=
\eta_c$ the solution becomes singular, with $C=0$. This shows that the 
flat cylindrical solution (\ref{cigar}) with $\eta=\eta_c$ is unstable 
with respect to the introduction of an arbitrarily small cosmological 
constant $\Lambda$.

\section{Solutions with $\Lambda<0$}

The solutions (\ref{s11}) and (\ref{s21}) given in the previous section 
also allow for negative values of $\Lambda$. 
There are actually three different possibilities, since now
$n-2-\kappa^2\eta^2$ can be either positive, negative, or zero.

\subsection{Class I}

For $\Lambda<0$ and depending on the sign of $n-2-\kappa^2\eta^2$,
we can define a new radial coordinate $\chi$ as
\beq
\xi=
 \sqrt{K}\alpha \sinh \chi,
     ~\sqrt{K} e^{\chi},
     ~\sqrt{K}\alpha \cosh \chi
\label{ch1}
\eeq
for $n-2-\kappa^2\eta^2$ less, equal and greater than zero, respectively.
The range for the new coordinate is
 $0\leq\chi<\infty$ in the first case and
$-\infty <\chi<\infty$ in the other two.
Then we can write the metric as
\beqa
 & \quad \hat{R}<0: \quad&
ds^2=K \left[ d\chi^2+ \alpha^2 \sinh^2 \chi
\;d\Omega_{n-1}^2   \right. \nn \\
&& \left.
+\cosh^2 \chi d{\hat s}_-^2\right],
\label{c12} \\
      & \quad \hat{R}=0: \quad &
ds^2=K \left[ d\chi^2+ e^{2 \chi}
(d\Omega_{n-1}^2+d\hat{s}_0^2) 
\right], \\
     & \quad \hat{R}>0: \quad &
ds^2=K \left[ d\chi^2+ \alpha^2 
\cosh^2 \chi
d\Omega_{n-1}^2+ \sinh^2\chi d\hat{s}_+^2 
\right].  \label{c14}
\eeqa

Here, $d{\hat s}^2_\pm$ is the metric on a space of constant
curvature satisfying (\ref{bm}) with ${\hat R}=\pm p(p-1)$, and
$d{\hat s}_0^2$ is a Ricci-flat metric.  In the
case of negative curvature, we can
choose, for example, the anti-de Sitter space
\beq
d{\hat s}_-^2=(-dt^2+\sin^2 t \;(d\psi^2+\sinh^2\psi
\;d\Omega_{p-2}^2)).
\eeq
Flat space metric can be used for
$d\hat{s}_0^2$, and the de Sitter metrics (\ref{ds1})
can be used for the constant positive curvature space $d{\hat
s}_+^2$. 

For ${\hat R}<0$, the defect is located at $\chi=0$.  For
${\hat R}=0$ it is removed to $\chi=-\infty$, and for ${\hat
R}>0$ there is no defect at all.  In the latter case, there is a minimum
radius for the $(n-1)$-spheres in
the extra dimensions, $r_{min}=K\alpha$.  We thus have a wormhole
connecting a monopole configuration at $\chi>0$ with an antimonopole
configuration at $\chi<0$. 

\subsection{Class II}

For the solutions defined by expressions (\ref{s21})
we find a similar situation. With a new
coordinate $\chi$ defined as in (\ref{ch1}), but with $\alpha$
replaced by ${\tilde\alpha}$, we have
\beqa
 & \quad \hat{R}>0: \quad&
ds^2=K \left[ d\chi^2+ {\tilde\alpha}^2 \sinh^2 \chi
(d\Omega_{n-1}^2+d\hat{s}_+^2) 
\right] 
\label{cl2}\\
      & \quad \hat{R}=0: \quad &
ds^2=K \left[ d\chi^2+ e^{2 \chi}
(d\Omega_{n-1}^2+d\hat{s}_0^2) 
\right] \\
     & \quad \hat{R}<0: \quad &
ds^2=K \left[ d\chi^2+ {\tilde\alpha}^2 \cosh^2 \chi
(d\Omega_{n-1}^2+d{\hat s}_-^2)
\right].
\eeqa
Once again, the metric (\ref{cl2}) is singular at $\chi=0$ even in the
absence of a defect $(\eta=0)$, and we dismiss this solution as
unphysical.

\subsection{Class III}

We finally consider the cylindrical metric {\it ansatz} (\ref{newansatz}).  
The solutions of Eqs.(\ref{cyl1})-(\ref{cyl3}) for $\Lambda<0$ have
the form
\beqa
 & \quad \hat{R}>0: \quad&
ds^2=C^2d\Omega_{n-1}^2+\omega^{-2}(d\chi^2+\sinh^2\chi d\hat{s}_+^2),
\label{ccl2}\\
     & \quad \hat{R}<0: \quad &
ds^2=C^2d\Omega_{n-1}^2+\omega^{-2}(d\chi^2+\cosh^2\chi d\hat{s}_-^2),
\label{ccl3}\\
     & \quad \hat{R}=0: \quad &
ds^2=C^2d\Omega_{n-1}^2+d\chi^2+e^{\pm2\omega\chi} d\hat{s}_0^2,
\label{ccl1}
\eeqa
where
\beqa
& \omega=\sqrt{{-2\Lambda\over p(n+p-2)}}, &\\
& C^2=-(n+p-2)(\kappa^2\eta^2-(n-2))/2\Lambda.&
\label{C}
\eeqa

Of greatest interest are the flat brane solutions (\ref{ccl1}) which
generalize the solutions considered by Gregory \cite{Gregory} 
in the $n=2$ case.
The geometry of the extra dimensions in the metric (\ref{ccl1}) is that
of a cylinder with a cross-section being an $(n-1)$-sphere of a fixed radius
$C$.  It would be interesting if this solution could be matched to an 
appropriate interior solution, so that the complete geometry is that
of a `cigar'.  Gregory \cite{Gregory} has argued that this is 
possible for $n=2$, but her analysis does not directly apply to $n\geq
3$.

Cigar-like defect solutions with an exponential warp factor 
would be of interest, since they would
have features similar to those of the Randall-Sundrum geometry.   
If the brane is located at $\chi=0$ and the asymptotic metric is given
by (\ref{ccl1}) with a negative sign in the exponential, 
then the volume of the extra
dimensions would be finite, despite their infinite extent in the
$\chi$-direction.  As in the Randall-Sundrum case, most of the
volume would be concentrated near the brane, and one can expect that
gravitons would be effectively confined to the brane.

The right-hand side of (\ref{C}) should be positive, so
we must have
$\kappa^2\eta^2-(n-2)>0$. While this does not give any additional 
information for $n=2$, this condition requires a super-Planckian
symmetry breaking scale, $\eta>\kappa^{-1}$, for the defects when $n>2$.

\section{Instanton solutions}

Euclidean continuations of brane-world solutions are of interest,
since they can be interpreted as gravitational instantons describing
quantum nucleation of a brane-world.  The nucleation probability is
given by
\beq
{\cal P}\propto e^{\pm |S|},
\label{calp}
\eeq
where $S$ is the instanton action.  The choice of sign in the
exponential is determined by the choice of boundary conditions for
the wave function of the universe.  The lower sign is chosen for the
tunneling and Linde boundary conditions, and the upper sign for the
Hartle-Hawking boundary condition \cite{QC}.  For definiteness we
shall adopt the tunneling boundary condition below.

For the instantons to give a nonvanishing
contribution to the nucleation probability, 
they must have a finite action, with instantons of the smallest
absolute value of the action
giving the dominant contribution.
The action is typically extremized for solutions of the highest
symmetry, so we shall consider instantons with $d{\hat s}^2_+$,
$d{\hat s}^2_-$ and $d{\hat s}^2_0$ being maximally symmetric spaces
of positive, negative and zero curvature, that is, Euclidean de
Sitter, anti-de Sitter, and flat spaces, respectively.

The Euclidean action for our model is given by
\beq
S=-{1\over 2\kappa^2}
\int d^{(n+p)}x \sqrt{-g}\; [R-2\Lambda -2\kappa^2 L(\phi)],
\eeq
where $R$ is the $D$-dimensional scalar curvature and 
$L(\phi)$ is the scalar field Lagrangian.  
We can eliminate $R$ by making use of Einstein's equations to obtain 
\beq
R=2 \kappa^2 L(\phi)+{2(n+p)\over n+p-2}\;\Lambda
\eeq
and
\beq
S=-{\Lambda\over{\kappa^2(n+p-2)}}\int d^{n+p}x\sqrt{-g}
\eeq

For class-I and class-II solutions with $\Lambda<0$, the volume of the
transverse space is infinite, and $|S|=\infty$.  If cigar-like
class-III solutions exist, they may have a finite transverse volume, but
the action is still infinite due to the divergence of the
$p$-dimensional volume of the flat brane worldsheet.  Hence, we only
need to consider solutions with $\Lambda>0$.  In this case the
curvature of the brane must be positive, ${\hat R}>0$, and thus the
metric $d{\hat s}^2$ should be that of a Euclidean de Sitter space,
that is, a $p$-sphere:
\beq
ds^2_E=K \left[ d\chi^2+ \alpha^2 \sin^2\chi
d\Omega_{n-1}^2+\cos^2 \chi 
(d\psi^2+\sin \psi^2 d\Omega_{p-1}^2)
\right].
\label{sE}
\eeq

One can model the nucleation of a closed universe 
with a brane 
by allowing
$\psi$ to vary in the interval $[0,\pi/2]$ in the
Euclidean region and then continuing it in the imaginary direction in the 
Lorentzian region,
$\psi=\pi/2+i t$.  This turns (\ref{sE}) into the 
metric
\beq
ds^2=K \left[ d\chi^2+ \alpha^2 \sin^2\chi
d\Omega_{n-1}^2+\cos^2 \chi
(-dt^2+\cosh^2 t \;d\Omega_{p-1}^2)
\right]
\eeq 
describing an expanding
braneworld.

We can easily calculate the action for the instanton 
solution (\ref{sE}): 
\beqa
S&=&{1 \over 2 \kappa^2}{V_{p}\over 2}\;V_{(n-1)} K^{(n+p)/2} 
\alpha^{n-1} \int_0^{\pi} d\chi  |\cos \chi|^p (\sin \chi)^{(n-1)}
{4\Lambda\over n+p-2}\\*[15pt]
&=&{4 \over \kappa^2}\;
{ \sqrt{\pi}^{(n+p+1)}\over \Gamma[(n+p-1)/2]}
K^{(n+p-2)/2}\alpha^{n-1}.
\label{action}
\eeqa
where $V_k$ stands for the volume of a k-sphere of unit radius, that is, 
$V_k=2 \pi^{(k+1)/2}/ \Gamma[(k+1)/2]$.

Apart from nucleation of the entire brane-world, the instanton
(\ref{sE}) can also describe nucleation of spherical branes in an
inflating $(n+p)$-dimensional de Sitter space.  The situation here is
very similar to the nucleation of circular loops of string and of
spherical domain walls in a $(3+1)$-dimensional de Sitter space, as
discussed by Basu {\it et. al.} \cite{Basu}.  The nucleation rate is
given by
\beq
\Gamma\propto e^{-B}
\eeq
with
\beq
B=S-S_0,
\eeq
where $S$ is the instanton action and $S_0$ is the action for the
Euclidean de Sitter space without a brane.  From Eq.(\ref{action}) we
have
\beq
B={4 \over \kappa^2}\;
{{\pi}^{(n+p+1)/2}\over \Gamma[(n+p-1)/2]}K^{(n+p-2)/2}(1-\alpha^{n-1}).
\eeq

The initial radius of the brane is $r=\sqrt{K}$.  After nucleation, it
is stretched by the exponential expansion of the universe.

\section{Conclusions}

In this paper we have found a number of solutions describing global
defects in a higher-dimensional space.  We assumed that the core of
the defect is centered on a $(p-1)$-dimensional brane and concentrated
on the case when the number of extra dimensions is $n\geq 3$.  

In the absence of a cosmological constant, we found that for all
$n\geq 3$ the defect
solution is very similar to that for a global monopole \cite{BV}.  The
brane worldsheet is flat, and there is a solid angle deficit in the
extra dimensions.  This is rather surprising, considering the fact
that solutions are very different for $n=1$ and $n=2$.  The maximal
solid angle deficit is reached at the critical value
$\eta_c=(n-2)^{1/2} \kappa^{-1}$, when the transverse metric becomes
that of a cylinder.

For a positive cosmological constant, $\Lambda >0$, 
our solutions describe spherical
branes in an inflating higher-dimensional universe.  In the limit
$\eta\to 0$, when the gravitational effect of the defect can be
neglected, the universe can be pictured as an expanding
$(p+n-1)$-dimensional sphere with a brane wrapped around it in the
form of a sphere of lower dimensionality $(p-1)$.  A nonzero $\eta$
introduces a deficit angle in the dimensions orthogonal to the brane
worldsheet.  It is interesting that the expansion rate of the universe
(and of the brane) is independent of the symmetry breaking scale
$\eta$ and is determined only by $\Lambda$, while the deficit angle is
determined by $\eta$ and independent of $\Lambda$.  Gravitational
instantons obtained by a Euclidean continuation of this class of
solutions have the geometry of a $(p+n)$-sphere with the brane
represented by a maximal $p$-sphere and with a deficit solid angle in
the dimensions transverse to the brane.  These instantons
can be interpreted as describing quantum nucleation either
of the entire brane-world, or of a spherical brane in an inflating
$(p+n-1)$-dimensional universe.  

Another class of solutions has curvature singularities even in the
absence of a defect $(\eta=0)$, and we have dismissed such solutions
as unphysical.

The third class of solutions has the geometry of a $(p+1)$-dimensional
de Sitter space, with the remaining $(n-1)$ dimensions having the
geometry of a cylinder.

We have also found 3 classes of solutions for $\Lambda<0$.  The first two
are essentially analytic continuations of the positive-$\Lambda$
solutions.  The third class is similar to Randall-Sundrum $(n=1)$ and
Gregory $(n=2)$ solutions, exhibiting an exponential warp factor.  If
solutions of the third class can be matched to appropriate interior
solutions in the defect core, one may be able to use them as a basis
for realistic brane-world models.

\section{Acknowlegements}

We are grateful to Gia Dvali, Jaume Garriga and Ruth Gregory  
for useful discussions and comments on the manuscript.  This work was
supported in part by the Basque Government
under fellowship number BFI.99.89
 (I.O.) and by the National Science
Foundation (A.V.).


\begin{thebibliography}{99}

\bibitem{Gia}
N. Arkani-Hamed, S. Dimopoulos and G. Dvali, Phys. Lett. {\bf B429},
263 (1998); Phys. Rev. {\bf D59}, 086004 (1999).
\bibitem{RS} L. Randall and R. Sundrum, Phys.Rev.Lett. {\bf 83}, 
  4690 (1999).
\bibitem{Sundrum} R. Sundrum, Phys.Rev. {\bf D59}, 085010 (1999).
\bibitem{CP} A. Chodos and E. Poppitz, Phys. Lett. B{\bf 471}, 119 (1999).
\bibitem{CK} A.G. Cohen and D.B.Kaplan, Phys.Lett. {\bf B470}, 52 (1999).
\bibitem{Gregory} R.Gregory, hep-th/9911015.
\bibitem{GS} J. Garriga and M. Sasaki, hep-th/9912118.
\bibitem{BV} M. Barriola and A. Vilenkin, Phys. Rev. Lett. {\bf 63},
341 (1989). 
\bibitem{Rama} 
A similar coordinate system in four dimensions has been used in
R. Basu and A. Vilenkin, Phys. Rev. {\bf D46}, 2345 (1992).
\bibitem{DG} O. Dando and R. Gregory, Phys. Rev. {\bf D58}, 023502 (1998).
\bibitem{RT}
I.H. Redmount and S. Takagi, Phys. Rev. {\bf D37}, 1443 (1988).

\bibitem{QC}
It should be noted that in the
case of the tunneling boundary condition Eq.(\ref{calp}) has been
shown to apply for homogeneous instantons and may need modification in
our case when the gradients of the fields are important.  Linde's wave
function was only defined for a simple minisuperspace model, and it is
not quite clear how to extend it to the present case.  For a
discussion of these issues and references to the literature, see,
e.g., A Vilenkin, Phys. Rev. {\bf D58}, 067301 (1998).
\bibitem{Basu}
R. Basu, A.H. Guth and A. Vilenkin, Phys. Rev. {\bf D44}, 340 (1991).

\end{thebibliography}
\end{document}